\documentclass[conference]{IEEEtran}
\IEEEoverridecommandlockouts
\usepackage{cite}
\usepackage{amsmath,amssymb,amsfonts}
\usepackage{algorithmic}
\usepackage{graphicx}
\usepackage{textcomp}
\usepackage{xcolor}
\usepackage{graphicx}
\usepackage[caption=false,font=footnotesize,labelfont=sf,textfont=sf]{subfig}

\def\BibTeX{{\rm B\kern-.05em{\sc i\kern-.025em b}\kern-.08em
    T\kern-.1667em\lower.7ex\hbox{E}\kern-.125emX}}
\begin{document}
\title{On the Activity Privacy of Blockchain for IoT }

\author{\IEEEauthorblockN{Ali Dorri$^{a,b}$, Clemence Roulin$^{a,c}$, Raja Jurdak$^{a}$, Salil S. Kanhere$^{b}$}
	\IEEEauthorblockA{\textit{$^{a}$DATA61, CSIRO, Australia. $^{b}$ CSE, UNSW, Sydney, Australia. $^{c}$ENAC}  \\
		(Ali.dorri,Salil.kanhere)@unsw.edu.au, cle.roulin@gmail.com,Raja.Jurdak@csiro.au }
	%\and
	\thanks{The first two authors contributed equally to this work.}
	%\IEEEauthorblockN{Ali Dorri}
	%\IEEEauthorblockA{\textit{CSE, UNSW}\\
	%	\textit{and   DATA61, CSIRO} \\
	%Sydney, Australia \\
	%ali.dorri@unsw.edu.au}
	%\and
	%\IEEEauthorblockN{Raja Jurdak}
	%\IEEEauthorblockA{\textit{DATA61, CSIRO} \\
	%Brisbane, Australia. \\
	%Raja.Jurdak@csiro.au}
	%\and
	
	%\IEEEauthorblockN{Salil S. Kanhere}
	%\IEEEauthorblockA{\textit{CSE, UNSW} \\
	%Sydney, Australia \\
	%Salil.kanhere@unsw.edu.au}
}

\maketitle

\begin{abstract}
	Security is one of the fundamental challenges in the Internet of Things (IoT) due to the heterogeneity and resource constraints of the IoT  devices. Device classification methods are employed  to enhance the security of IoT by detecting unregistered devices or traffic patterns. In recent years, blockchain has received tremendous attention as a distributed trustless platform to enhance the security  of IoT. Conventional device identification methods are not directly applicable in blockchain-based IoT as  network layer packets are not stored in the blockchain. Moreover, the transactions are  broadcast and thus have no destination IP address and contain a public key as the user identity, and  are stored permanently in blockchain which can be read by any entity in the network. We show that  device identification in blockchain introduces privacy risks as the malicious nodes can identify users' activity pattern by analyzing the temporal pattern of their transactions  in the blockchain. We study the likelihood of classifying IoT devices by analyzing their information stored in the blockchain, which to the best of our knowledge, is the first work of its kind. We use a smart home as a representative IoT scenario. First, a blockchain  is populated according to a real-world smart home traffic dataset.  We then apply machine learning algorithms on the data stored in the blockchain  to analyze the success rate of device classification, modeling both an informed and a blind attacker. Our results demonstrate success rates over 90\% in classifying devices. We propose three timestamp obfuscation methods, namely combining multiple packets into a single transaction, merging ledgers of multiple devices, and randomly delaying transactions, to reduce the success rate in classifying devices.  The proposed timestamp obfuscation methods can reduce the classification success rates  to as low as 20\%.
	
\end{abstract}


\section{Introduction}
The Internet of Things (IoT) brings connectivity to everyday devices and enables a wide range of personalized services to end users.  There have been many attacks reported on the IoT devices in recent years \cite{HackedIoTDevices,realIoTHack}. Conventional security solutions are too heavy weight for IoT and thus significant research has focused on improving IoT security \cite{jing2014security} \par 
\subsubsection{Overview}
Due to the ever increasing  number of IoT devices, it is of great importance for the operators of the smart environments, e.g., a smart campus, to determine the type of devices connected to their site and ascertain that the devices are functioning normally. Device classification methods are employed to address the aforementioned challenge. The authors in  \cite{feng2018acquisitional}  analyzed the real-time traffic of  IoT devices, which includes network layer packets, e.g., SMTP, to identify the semantic type of the device (e.g., a camera,  a motion sensor, or a smart light). In \cite{miettinen2017iot}, the authors employed software defined networking  to identify the devices that lack proper security configurations. The authors in \cite{sivanathan2018classifying} employed machine learning algorithms for device classification.  \par 

Blockchain has attracted tremendous attention as a promising approach to mitigate security risks in IoT due to its salient features which includes transparency, immutability, and decentralization. All transactions, i.e., communications between devices, are stored permanently in the blockchain.  Each block includes the hash of its previous block in the ledger, which  ensures immutability of the ledger. The modification  or removal of the block content, i.e.,  transactions, is impossible, since the hash maintained in the subsequent block will not match with the hash of the modified block.  \par 

The transactions are cryptographically sealed using public/private keys. The Public Key (PK) used in each transaction is employed as the identity of the transaction generator. This introduces a level of anonymity for the blockchain users as their real identity remains unknown to the participating nodes. To enhance their anonymity, the users may change their PK for each new transaction as in Bitcoin \cite{nakamoto2008bitcoin}. This protects  users against linking attacks, where malicious nodes  attempt to deanonymize a user by tracking multiple identities of the user.\par 
\subsubsection{Challenges}
Unlike conventional device identification methods where the main aim is to detect malicious devices, in blockchain-based IoT, device classification  is a privacy risk, where the goal is to expose the  user's activity patterns based on the sensed data. An attacker with the intention of unveiling a user's activities must first determine the type of sensing devices in the user's premises. The combination of user deanonymisation and device identification can therefore be a powerful tool for an attacker to determine a user's identity and activities. For example, in energy trading, the malicious nodes can identify the energy consumption pattern of the users, or in a smart city, the malicious node can link multiple transactions  to   track the historical record of the location of a particular vehicle. Conceptually, device classification in blockchain is similar to the linking attack \cite{dorri2019mof} where the attacker attempts to link multiple anonymous transactions to a particular user (see Section \ref{sec:lit-review}).\par 
The use of blockchain for recording IoT device transactions changes the design space for device classification due to the following reasons:  \par 
i)  Transactions are permanently stored in the blockchain which potentially creates  a large database of  historical  interactions of  devices. This potentially exposes more data  from the users to the attacker over a longer period.\par 
ii) In conventional IoT classification methods  the IP address of the source and destination entities can be retrieved from the packets which can be used  as input to assist in identifying the type of the devices, e.g., a device that frequently sends packets  to Nest servers  may potentially  be a smart thermostat. In contrast, blockchain transactions contain the PK as the identity of the involved entities and are broadcast to the network.  Blockchain participants can change their PK in each transaction which in turn introduces a level of anonymity.  \par 
iii) In conventional methods for IoT device identification, the identifier eavesdrops on real-time network traffic and thus should have physical access to the network where the IoT devices are deployed. The real-time traffic potentially contains network packets as well as exchanged data and communications between parties. The pattern of network management packets or the size of the packets can assist  in identifying the device. In contrast, in blockchain any entity  can launch   device identification  given that all participants can read the blockchain independent of the physical location of the entities. In blockchain, only the transaction corresponding to communications between devices which contains the  hash of the exchanged data can be accessed. The network layer packets are not stored in the blockchain, which in turn make it impossible for the identifier to identify a device based on such data. \par

\subsubsection{Contributions}
In this paper, we study the possibility of  device identification in IoT-based blockchain by analyzing the  patterns of recorded transactions in the blockchain. As an example,  a  Samsung camera that stores transactions in blockchain can be identified as a ``camera" by analyzing the pattern of its transactions. To the best of our knowledge, this is the first attempt to identify  device types in an IoT blockchain context.  In an IoT setting, each user owns a number of devices that collect and share  data with Service Providers (SP) and/or other users to offer personalized services to the user. Exposure of the  user's activity patterns results in serious privacy and security concerns, e.g., the attacker can infer the hours that a home is occupied by monitoring the temporal patterns of transactions generated by motion sensors. In most blockchain instantiations, the data of the IoT devices are not stored in the blockchain, but rather off-the-chain in a separate cloud storage with only the hash of the data being stored in the blockchain.  It is not necessary for the attacker to access the data to expose the user's activity as attackers can do so by monitoring the pattern of stored  communications, i.e., transactions, of IoT devices \cite{apthorpe2017smart}.  \par 

To study the device classification problem, we rely on  a  smart home as a representative case study. We use the smart home traffic dataset available in \cite{UNSWIoTDataset} and populate a blockchain by generating transactions corresponding to  device communications. We study  the success rate of  classification of the IoT devices, in terms of semantic type, based on transaction  patterns. To protect against this attack, we propose multiple timestamp obfuscation methods, such as combining multiple packets into a transactions, merging ledgers of multiple devices, and randomly delaying transactions, and study the impact of the proposed methods on attack success.\par 

The attacker's aim is to match temporal signatures observed in the transaction ledger to known inter-packet times of  devices in order to identify the type of devices.  As the attack is a pattern matching problem, we use machine learning, particularly decision trees, to model the attacker. We choose decision trees as our machine learning algorithm as the resulting classifications are more explainable and they work well for categorical data, which fits well for classifying different sensor types.  Our results show that attacks can correctly classify over 90\%  of devices in the ledger, while the timestamp obfuscation methods can reduce attack success rates to between 20-30\%.\par 
\subsubsection{Paper Structure}
The rest of the paper is organized as follows. Section \ref{sec:lit-review} discusses the related works. Details of  activity privacy in blockchain is outlined in Section \ref{sec:blockchain-for-anonymity}. Evaluation results are presented in Section \ref{sec-evaluation}. Section \ref{sec:discussions} presents discussions and concludes the paper.  \par

\section{Related Work}\label{sec:lit-review}
Blockchain users employ changeable PKs as their identity  to protect their transactions from being linked together, i.e., classified, that eventually may lead to user deanonymization. Studies show user deanonymization is still possible using blockchain and off-the-chain information \cite{ermilov2017automatic}, i.e., other publicly available information in the Internet. While existing studies on  blockchain transaction classification focus on Bitcoin, their findings are equally relevant to other cryptocurrencies that commonly record the sequence of coin exchange. \par 
The authors in \cite{conti2018survey}  argued that the analysis of the blockchain transaction pattern  involves three main steps:\par 
1) Transaction graph: the flow of the transactions in the blockchain can be represented as a graph. In this model, the transactions serve as nodes and  the PKs used as input/output of the transactions  serve as undirected edges between the nodes.  \par 
2) Address graph: this graph contains the flow of payments between multiple PKs and is inferred by analyzing the transaction graph. \par 
3) User graph: this graph contains the group of PKs that might belong to the same user based on the information in the address graph and heuristics available from Bitcoin. We discuss examples of Bitcoin heuristics below. \par 

The authors in \cite{androulaki2013evaluating}  considered two heuristics to create the user graph: multiple input transactions and shadow addresses. With multiple input transactions, a Bitcoin transaction can spend the output  of multiple transactions using multi-input transactions. The generator of the multi-input transaction must have the private keys corresponding to the output PK of the transactions to be spent. Thus, it is safe to assume that all these transactions belong to the same user. With shadow addresses, the Bitcoin user pays the remaining balance of his account to a new PK, i.e., changes his PK, in each transaction to protect his privacy. Accordingly, if the outputs of a  transaction are different from the inputs, one of the output addresses might be the shadow address of the same user.\par 

The authors in \cite{ermilov2017automatic} cluster blockchain addresses based on not only the blockchain transactions, but also the available off-the-chain data, e.g., the instances where the PK is mentioned along with a tag that can be a company name. The numerical results show that the proposed method can successfully classify the transactions with higher rate compared to methods that are only based on the blockchain information. \par 
Mixing services can be used to enhance user privacy \cite{bonneau2014mixcoin}. A central node, known as mixer, changes the identity of the user coin with a randomly chosen identity to break the link between the users identities and thus protect their privacy. However,  classifying multiple transactions of a user is still possible as proved by the authors in \cite{hong2018practical}. In \cite{hong2018practical} the authors deanonymized users  by exploiting the static and dynamic parameters used by the mixing service to the mix keys. \par 

%Most of the existing works focus on data available in the blockchain or off-the-chain to deanonymize a user. In \cite{fanti2017deanonymization} the authors analyzed Bitcoin anonymity based on  network level information, i.e., IP addresses. To distribute transactions, Bitcoin originally employed trickle spreading protocol, which is a gossip based protocol. However, this protocol was shown to be vulnerable to user deanonymization and was thus  replaced by diffusion spreading protocol in 2015. The authors in \cite{fanti2017deanonymization} studied the anonymity of both these protocols. The authors modeled the user deanonymization mathematically as inferring the source of a random spreading in a tree. Mathematical equations prove that the source of spreading can be identified in both new protocols, which in turn proves vulnerability of these protocols  against user  deanonymization.   \par 

The classification of transactions of the blockchain is largely studied in Bitcoin. As blockchain is being widely applied to non-monetary applications, e.g., IoT, it is critical to study the anonymity of the blockchain in such systems.  To the best of our knowledge, this work is the first to analyze the vulnerability of blockchain to classify device identification in  IoT. Device classification in blockchain-based IoT leads to further privacy concerns as the classifier can identify the transactions pattern which potentially can be linked to a user and reveal the users activities. Thus, in the rest of the paper we consider device classification for IoT as an attack against user anonymity.   \par  

\section{Activity privacy of Blockchain in IoT} \label{sec:blockchain-for-anonymity}
In this section we discuss the implementation setting, attack model, and different timestamp obfuscation methods to study the device type identification risk in blockchain-based IoT. 

\subsection{Blockchain setup}\label{sec-implementation}
\begin{figure}
	\centerline{\includegraphics[width=9cm,height=7cm,keepaspectratio]{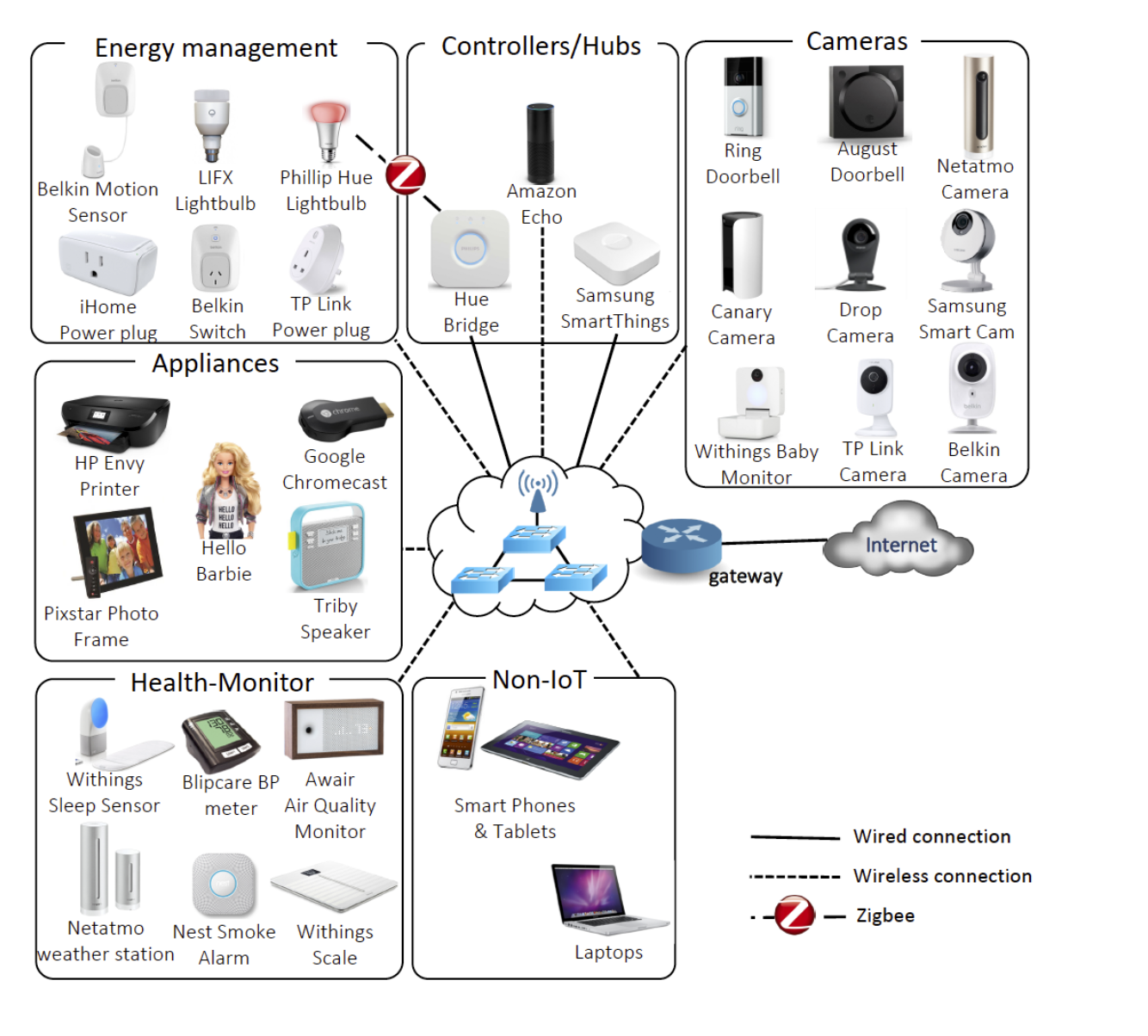}}
	\caption{ The IoT network testbed from  \cite{UNSWIoTDataset}.}
	\label{fig:overal-smart home}
\end{figure}

We first describe the real-world smart home network traffic data set used in our work, followed by a discussion about the blockchain implementation. We use the smart home traffic dataset available at \cite{UNSWIoTDataset} that is gathered from a real-world deployment of smart home IoT devices. The dataset contains the network traffic of 30 IoT devices for a period of two weeks. Figure \ref{fig:overal-smart home} depicts the network setup and types of IoT devices in this dataset. \par 

The blockchain is populated with transactions corresponding to the communications of IoT devices in the dataset. For each communication that a device makes with other devices or nodes, e.g.,  Service Providers (SP), a transaction is generated and stored in the blockchain.  It is assumed that devices change their PK for each transaction generated in the network, which potentially achieves a level of anonymity.  To focus on the vulnerability of stored  transactions to sensor type classification, we intentionally abstract out the following aspects of the network traffic and the blockchain: (a) the network management traffic, e.g., SMTP, from the dataset, as we assume that attackers only have access to the stored ledger rather than real-time network traffic; (b) the consensus algorithm, as the temporal patterns of the transactions are independent of the consensus algorithm. The structure of each transaction is as follows:\par 
$ T\textsubscript{ID} || P.T\textsubscript{ID} || timestamp || Output || PK || Sign $\par 
Where \textit{T\textsubscript{ID} } is the unique identifier of the transaction which is the hash of the transaction content. The transactions belonging to the same node need to be chained together to prevent Sybil attacks where a malicious node pretends to be multiple nodes by generating fake transactions  \cite{dorri2017lsb} (detailed in the rest of this section). \textit{P.T\textsubscript{ID}} is the identity of the previous transaction in the same ledger.   \textit{timestamp} is the time that each transaction is generated and corresponds to the time that the packets are generated in the database. \textit{Output} is the hash of the PK that the device will use in the next transaction.  The last two fields are the PK of the transaction generator and its corresponding signature. If the communication corresponding to the transaction involves data, the transaction generator will sign the hash of the data and populate it in \textit{sign} field. Otherwise, the hash of the transaction is signed. Note that in most existing blockchains, the data of IoT devices is not stored in the blockchain due to packet and processing overheads. Instead, only the hash of the data is stored which can still ensure data integrity.\par 

A single chain may  be utilized by multiple devices to enhance the anonymity of the user. In this case, it is difficult for the malicious node to link the transactions in the ledger to each device. The first transaction in a ledger is called \textit{genesis transaction}. The process of generating a genesis transaction varies based on the specific instantiation of the blockchain. However, all of these methods include a mechanism to limit the number of genesis transactions a user can generate, e.g., burning coins in Bitcoin  \cite{dorri2017lsb}. This essentially protects against Sybil attacks.  \par 
In our implementation, a single node acts as miner and collects all transactions generated by the devices. Once the number of transactions reaches a pre-defined value, known as \textit{blocksize}, the miner forms a new block and appends it to the blockchain. The transactions follow the same timestamps as the communications in the dataset. Once the blockchain is populated, we apply machine learning algorithms to identify devices based on different attack models.

\subsection{Attack Models} \label{subsec:attack-model}
As shown in  Table 1, the pattern of transactions mostly represent a sequence of in-order numbers. Different patterns share some features, e.g., a separation of 0.207s is found for both the Smarts Things and the Nest smoke alarm. This pattern can be best represented by trees, thus, the machine learning algorithms, employed by the attacker, use decision trees to analyze the pattern of transactions in the blockchain and classify devices. The attacker can read the stored transactions and blocks in the blockchain, but cannot decrypt the data associated with the transactions without the corresponding private key.  The attacker first trains the machine learning algorithm on a local network, referred to as testnet in the rest of the paper. Depending on the testnet, the ability of the attacker to detect the devices varies. We study the following two attack models:
\begin{itemize}
	\item Informed attack: In this attack, the attacker can  determine  the number and exact type (manufacturer) of devices used in a smart home, and aims to map its known device list to specific PKs in order to infer the user's activities. We model an informed attacker by using a 10-fold cross validation analysis, where the training process always ensures that the entire range of sensor devices in the home are represented in the trained algorithms. As  the attacker  knows the type of devices in the smart home, he can  acquire similar devices and collect sufficient  network traffic data to populate a comprehensive training set. While this attack is unlikely in practical scenarios, as it  requires the attackers to monitor the user's sensor acquisitions over a period of time prior to launching the attack, we include it as a worst case scenario for activity privacy risk analysis.
	
	\item Blind attack: In this attack, the attacker does not know the number and type of devices installed in the smart home. We model blind attackers by using training data that is collected from a small scale IoT network containing a few popular IoT devices.  We then use  test sets which contain fewer or more types of devices than in the training set. For instance, a motion sensor might be in the test set but no in the training set. The blind attacker can recreate a lab of his own with some typical smart home devices, and train his machine learning algorithm with the data from this lab.  The training set may contain all, some, or even none of the devices in the target smart home as the attacker does not know the type and number of devices installed in the target smart home. 
\end{itemize}

\subsection{Timestamp Obfuscation}\label{subsec:anonymity-level}
\begin{table}
	
	\begin{tabular}{|c|c|} \hline
		Device & Patterns of frequent time separation (in s)\\ \hline
		Smart\_Things & [0.207,58,0.207,58,…]\\ \hline
		Amazon\_Echo & [0.217,30,0.004,30,…]\\ \hline
		TPLink\_Camera & [0.12,61,0.12,61,…]\\ \hline
		Samsung\_Camera & [0.165,30,0.165,30,…]\\ \hline
		Drop\_Camera & 1.03 or 0.2\\ \hline
		Insteon\_Camera2	& [9x$<$0.0001,0.216,300,...]\\ \hline
		Baby\_Monitor & [600,0.28,600,0.28,…]\\ \hline
		TPLink\_Smartplug & [0.24,236,0.24,236,…]\\ \hline
		TPLink\_Smartplug & [0.12,236,0.12,236,…]\\ \hline
		iHome & [60,0.205,60,0.205]\\ \hline
		Nest\_Smockalarm	& [0.207,0.015,0.207,0.015,…]\\ \hline
		Netatmo\_Weather	& [1.72,0.33,1.72,0.33,…]\\ \hline
		Sleep\_Sensor & [10,0.276,10,0.276,…]\\ \hline
		Lifx\_Smartbulb & 1.92 or 60\\ \hline
		Triby\_Speaker & [120–0.3-120–0.3-56–0.3,…]\\ \hline
		Pix\_Photoframe & [0.31 or $>$=0.3,65,650]\\ \hline
		HP\_Printer & 90\\ \hline
	\end{tabular}
	\label{tab:devices}\caption{Inter-packet temporal patterns for devices}
\end{table}
In this section, we discuss different timestamp obfuscation methods employed for increasing resilience to device classification in blockchain. The  attacker can classify the transactions using their timestamp.  Recall that only the hash of the exchanged data  is stored in the transactions. By looking at the timestamp of   the transactions for  each device, the attacker can easily identify patterns for more than half the devices of our sample (see Table~\ref{tab:devices}). These patterns make it possible to classify observed transactions and identify which device generated them.

We define three timestamp obfuscation methods that delay transaction generation, combine packets into a single transaction, or combine transactions from multiple devices into a single ledger, and compare them with a baseline approach that does not employ timestamp obfuscation. In the baseline approach,   a transaction is generated for each single packet in the dataset of a device. Our first obfuscation method introduces a random delay for each transaction. More specifically, the transaction corresponding to communication `\textit{c}' is generated within the period of  [t\textsubscript{c}, t\textsubscript{c+ MaxDelay} ], where t\textsubscript{c} is the time that a communication occurs, and \textit{MaxDelay} is the maximum possible delay defined by the user. 	This method changes the pattern of transactions stored in the blockchain and increases the complexity of classifying transactions.
The random delay is generated independently for each transaction, so the transaction ordering may also change. \par 
\begin{figure*} [h]	% h = means that exaclty here
	\begin{center}
		%\subfloat[]{\label{fig:cloud-storage-pic} \includegraphics[width=7cm,height=7cm,keepaspectratio]{pics/storage.jpeg} }\hfill
		%	\mbox{
		%\subfloat[]{\label{fig:informed-delay} \includegraphics[width=8cm,height=7cm,keepaspectratio]{pic/informed-delay.jpg} }
		\subfloat[]{\label{fig:informed-delay-day}%\vspace{-2cm}
			\includegraphics[width=8cm]{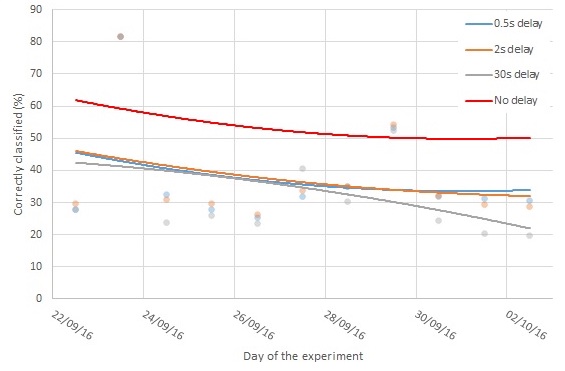} }
		%\subfloat[]{\label{fig:blind-delay} \includegraphics[width=8cm,height=7cm,keepaspectratio]{pic/blind-delay.jpg} }			
		\subfloat[]{\label{fig:blind-delay-day}
			\includegraphics[width=8cm]{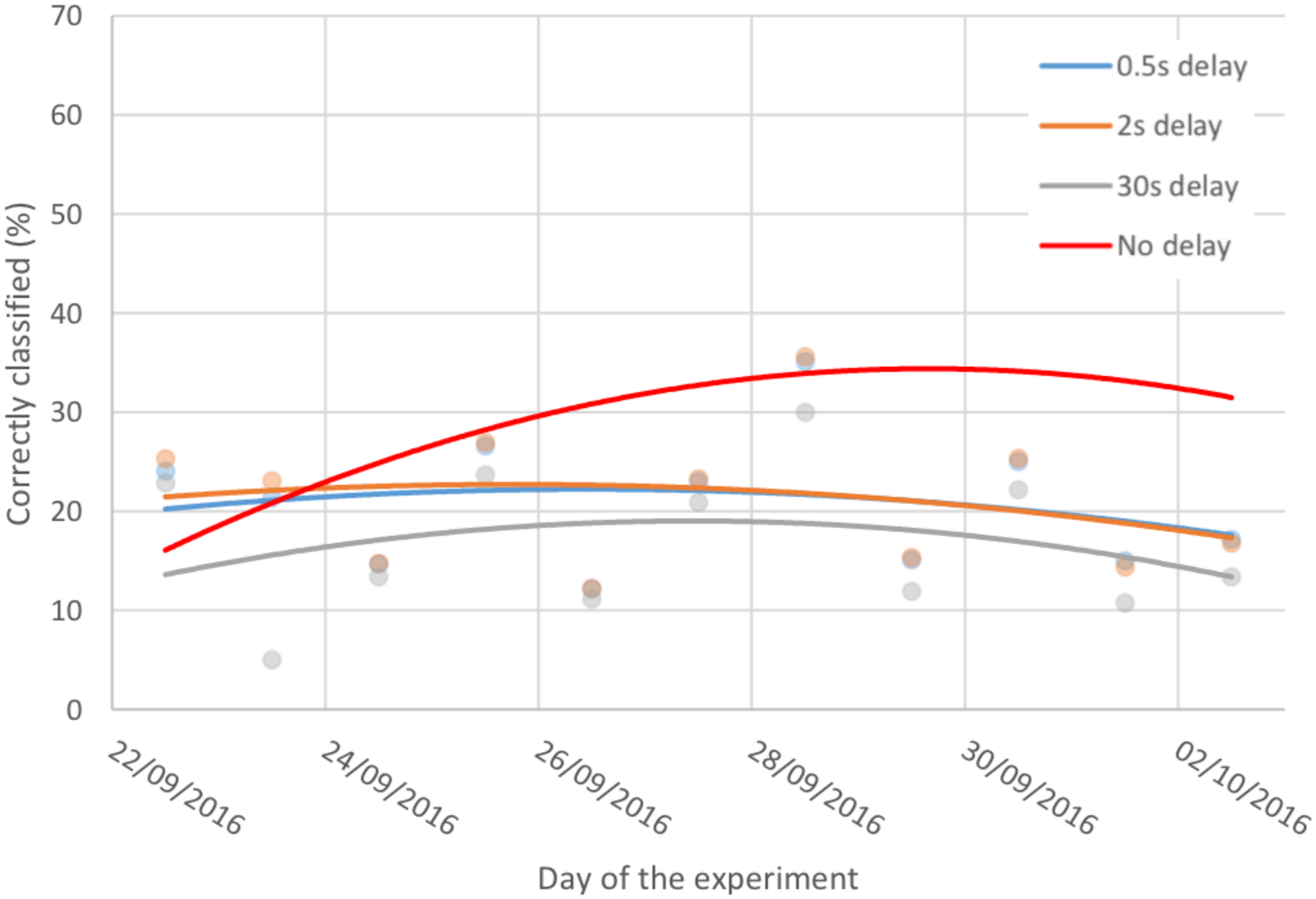} }
		%}
		%	\vspace{-3cm}
		\caption{The impact of delayed transactions for informed attacks (a) and blind attacks (b)}	\label{fig:delayed}
		%\label{fig:combination-localIL-and-transactions}
	\end{center}
	
\end{figure*}

In the baseline implementation  all  transactions of the same source (i.e., of a same device) are assigned to the same ledger. The distribution of the devices amongst the ledgers has a significant impact on the overall privacy as it links different transactions together. If all transactions of the same device are chained together, changing the PK per transaction will not affect the anonymity of the user as clearly all transactions in a ledger belong to the same user (no matter how the PK changes). Thus, in the second  proposed obfuscation method, called multi-node ledgers, a single ledger is shared for storing transactions among multiple devices. This potentially protects against  the attacker that evaluates the transactions of a ledger  to infer the pattern of transactions and thus contributes to safeguarding device types. \par 

IoT devices sometimes send data in bulk, for instance, when reporting highly relevant events. Large data payloads are  fragmented into multiple packets that are transmitted within a short timeframe. An attacker can exploit payload fragmentation by observing the stream of packets to infer the device type. Therefore, our third measure for data obfuscation creates multi-packet transactions, where we combine multiple packets from the same sender into one transaction. This can also be considered as summarizing all the communications in a single transaction. This timestamp obfuscation method reduces the likelihood of device type classification as: i) the volume of available data for the attacker decreases, ii) the pattern of transactions will not match with pattern of communications by the device which increases the difficulty of classifying the transactions. 

\section{Evaluation} \label{sec-evaluation}
In this section, we evaluate our three timestamp obfuscation methods through experiments on the empirical smart home dataset. We evaluate these methods for both informed and blind attack. For each measure, we synthetically create transactions based on the packet traces and then evaluate an attacker's ability to correctly classify the devices in the trace. We first evaluate these measures individually, then combine them to understand their cumulative benefits.
\subsection{Delayed transactions}
Considering the observed characteristic separation times observed in the devices' patterns (Table~\ref{tab:devices}), we introduce random delays generated in intervals of \([0, 0.5]\), \([0, 2]\) and \([0, 30]\) seconds respectively. The time separating two successive transactions of the same device is less regular, and the previously identified patterns are at least partly disturbed. \newline

\begin{figure} 	% h = means that exaclty here
	\vspace{-4cm}
	\includegraphics[width=8cm]{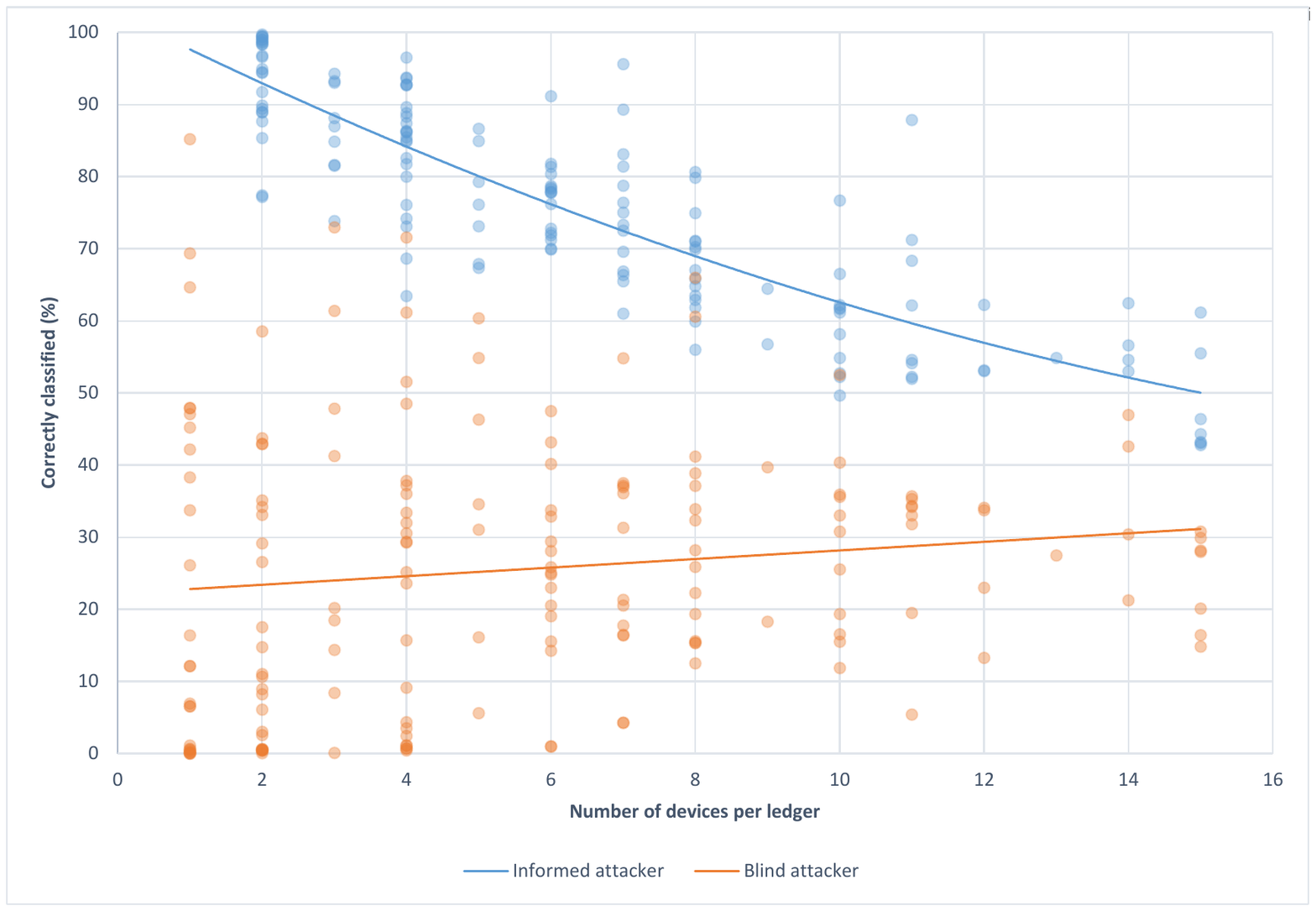} 
	%\subfloat[]{\label{fig:cloud-storage-pic} \includegraphics[width=7cm,height=7cm,keepaspectratio]{pics/storage.jpeg} }\hfill
	%\subfloat[]{\label{fig:informed-devices} \includegraphics[width=9cm]{pic/pictures/Singlepacket_Multinode.pdf} }
	%\subfloat[]{\label{fig:blind-devices} \includegraphics[width=9cm]{pic/pictures/Multinode_SinglevsMultiPacketTx_Blind.pdf} }
	\vspace{-3cm}
	\caption{ The impact of multi-node ledgers in informed and blind attack models.}
	%\label{fig:combination-localIL-and-transactions}
	\label{fig:multinode}
	
\end{figure}

\begin{figure*} 	% h = means that exaclty here
	\begin{center}
		%\subfloat[]{\label{fig:cloud-storage-pic} \includegraphics[width=7cm,height=7cm,keepaspectratio]{pics/storage.jpeg} }\hfill
		%	\mbox{
		%\subfloat[]{\label{fig:informed-delay} \includegraphics[width=8cm,height=7cm,keepaspectratio]{pic/informed-delay.jpg} }
		\subfloat[]{\label{fig:informed-delay-day} \includegraphics[width=8cm,keepaspectratio]{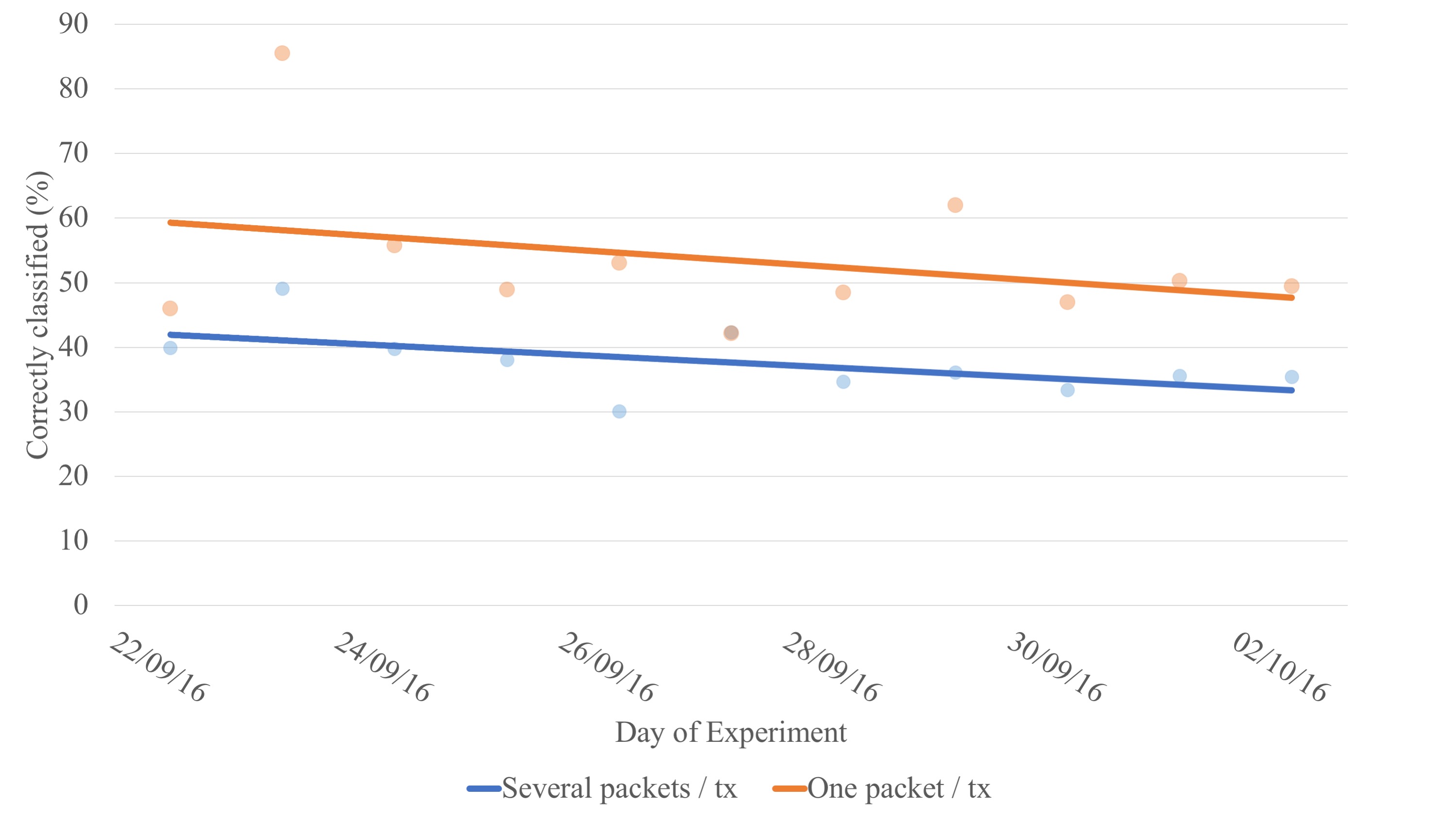} }
		%\subfloat[]{\label{fig:blind-delay} \includegraphics[width=8cm,height=7cm,keepaspectratio]{pic/blind-delay.jpg} }			
		\subfloat[]{\label{fig:blind-delay-day} \includegraphics[width=8cm,keepaspectratio]{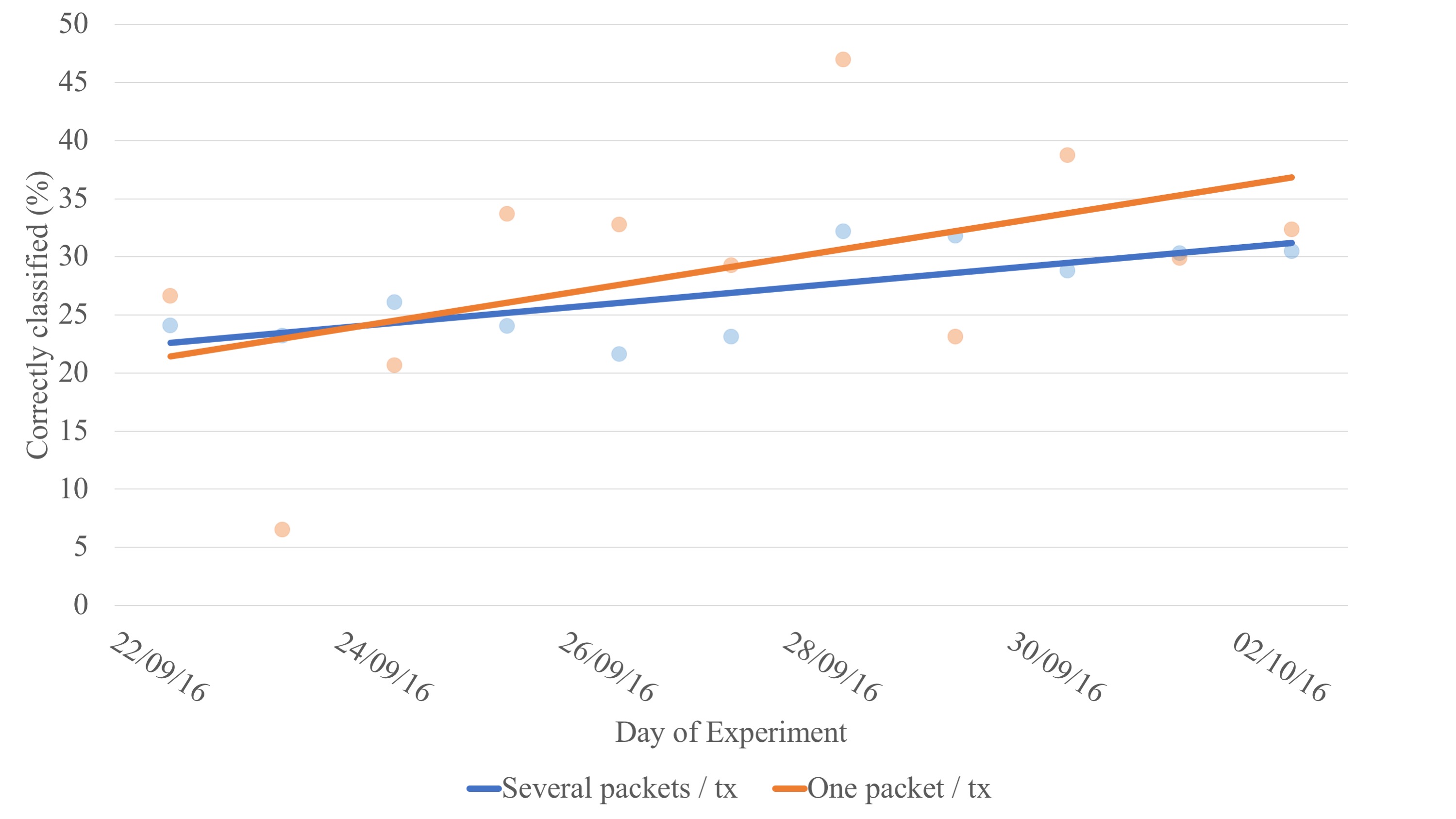} }
		%}
		\caption{The impact of multi-packet transactions for informed attacks (a) and blind attacks (b)}	\label{fig:multipacket}
		%\label{fig:combination-localIL-and-transactions}
	\end{center}
	
\end{figure*}

Figure~\ref{fig:delayed} shows the results for delayed transactions, for both informed  and blind attacks. Applying any of these random delays decreases the accuracy of the classification by more than 15\% for the informed attacker model. There is only a small difference between the results obtained for experiments with the  \([0, 0.5]\) and \([0, 2]\) intervals. This can be explained by the fact that examining the devices' patterns, the recognisable times are either \(<0.4\) seconds or \(\geq 28\) seconds. Testing two different delays in this interval has close results, as it wouldn't disturb the patterns. A delay of about 30 seconds creates a significant step of 30 seconds of separation, and makes it harder to identify patterns and to classify the transactions. For blind attacks, the classification accuracy drops for all approaches.  However, there is still a reduction of over 10\% with delayed transactions relative to the baseline approach. 

\subsection{Multi-device ledgers}
To see the impact of the number of devices per ledger, we randomly assign devices to a common ledger and  classify their combined transactions, while varying the number of devices per ledger. Figure~\ref{fig:multinode} shows the results. In the informed attack model, multi-node ledgers significantly reduce attack success rates from nearly 98\% for single node ledgers to around 50\% for ledgers with 17 nodes. When a small number of devices are pooled together, it is easier to differentiate the transactions' origins when we already know their behaviour, as they each have their own distinct temporal patterns. In contrast, including transactions from a larger number of devices in a ledger convolves inter-transaction times across devices, which is further accentuated when two devices have similar  transaction temporal patterns. It also becomes harder to infer patterns based on the separation times, as two back-to-back transactions are less likely to originate from the same device if we increase the number of devices included (and therefore the total number of transactions). 
\begin{figure*} 	% h = means that exaclty here
	
	\begin{center}
		%}
		%\subfloat[]{\label{fig:cloud-storage-pic} \includegraphics[width=7cm,height=7cm,keepaspectratio]{pics/storage.jpeg} }\hfill
		%	\mbox{
		%\subfloat[]{\label{fig:informed-delay} \includegraphics[width=8cm,height=7cm,keepaspectratio]{pic/informed-delay.jpg} }
		\subfloat[]{\label{fig:informed-delay-day} \includegraphics[width=9cm,keepaspectratio]{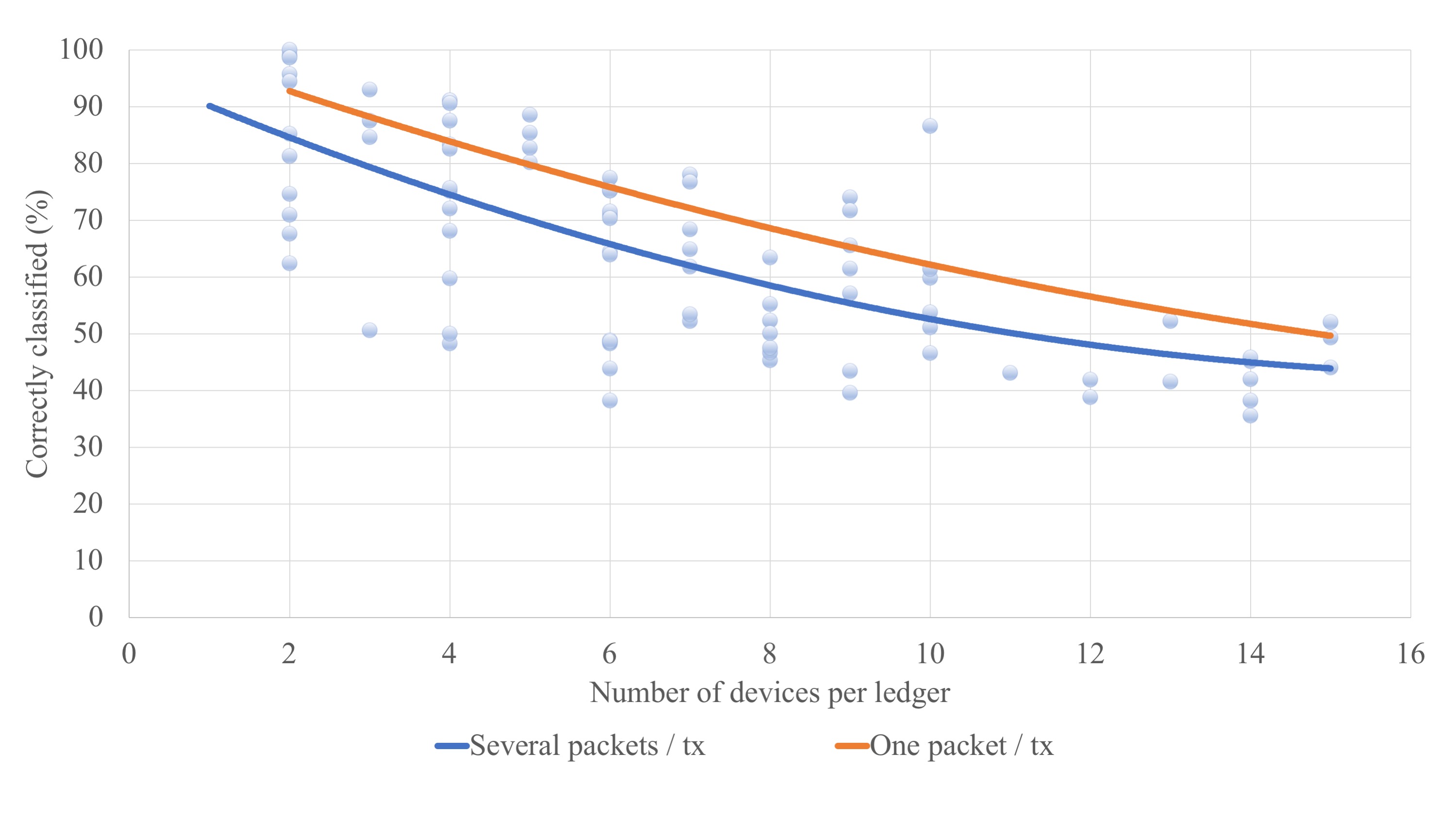} }
		%\subfloat[]{\label{fig:blind-delay} \includegraphics[width=8cm,height=7cm,keepaspectratio]{pic/blind-delay.jpg} }			
		\subfloat[]{\label{fig:blind-delay-day} \includegraphics[width=9cm,keepaspectratio]{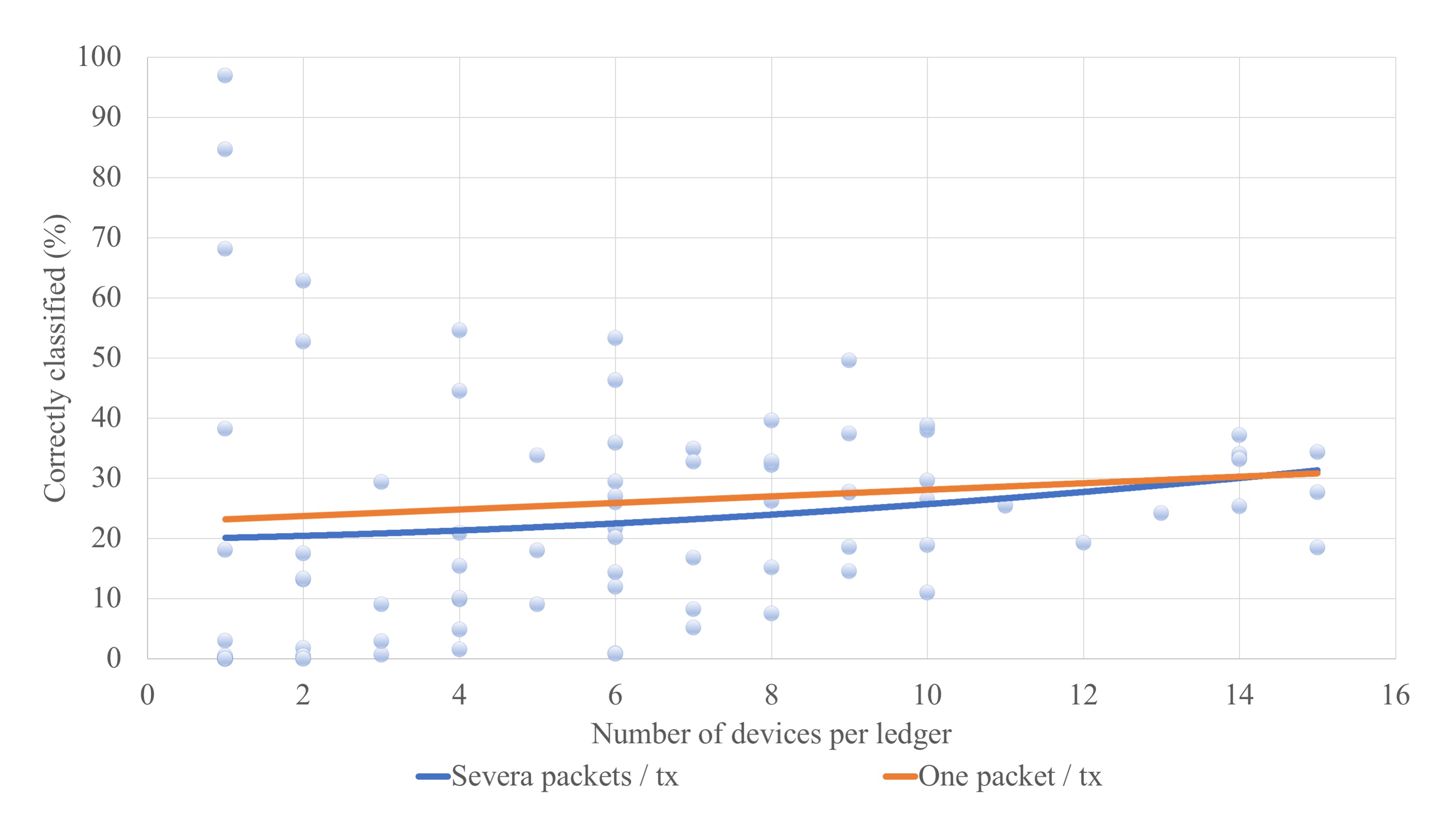} }
		\caption{The combined impact of multi-packet transactions and multi-node ledgers for informed attacks (a) and blind attacks (b)}\label{fig:multipacketledger}
		%\label{fig:combination-localIL-and-transactions}
		
	\end{center}
	
\end{figure*}

The overall success rates for blind attacks is considerably lower, and interestingly, the general trend for average attack success rate slightly increases with more devices per ledger. In this attack model, we  train our algorithm to recognize an available subset of all possible devices, hoping that at least devices present in the home (the test) are in our training set. The test set can therefore  vary significantly from our training set. A large number of devices within the same ledger will make the test set more similar to the training set, so the classification accuracy increases on a large sample. For smaller numbers of devices inside a ledger, we observe a high variance in the results. In fact, the trend in reduced variance with an increase in the number of devices per ledger is evident. Given this high variance across trials, we examine the maximum possible attack success in the model. Increasing the number of devices per ledger from 1 to 17 reduces the maximum attack success rate from around 85\% to 30\% for blind attacks.

\subsection{Multi-packet transactions}
\begin{figure*} 	% h = means that exaclty here
	\begin{center}
		%\subfloat[]{\label{fig:cloud-storage-pic} \includegraphics[width=7cm,height=7cm,keepaspectratio]{pics/storage.jpeg} }\hfill
		%	\mbox{
		%\subfloat[]{\label{fig:informed-delay} \includegraphics[width=8cm,height=7cm,keepaspectratio]{pic/informed-delay.jpg} }
		\subfloat[]{\label{fig:informed-delay-day} \includegraphics[width=9cm,keepaspectratio]{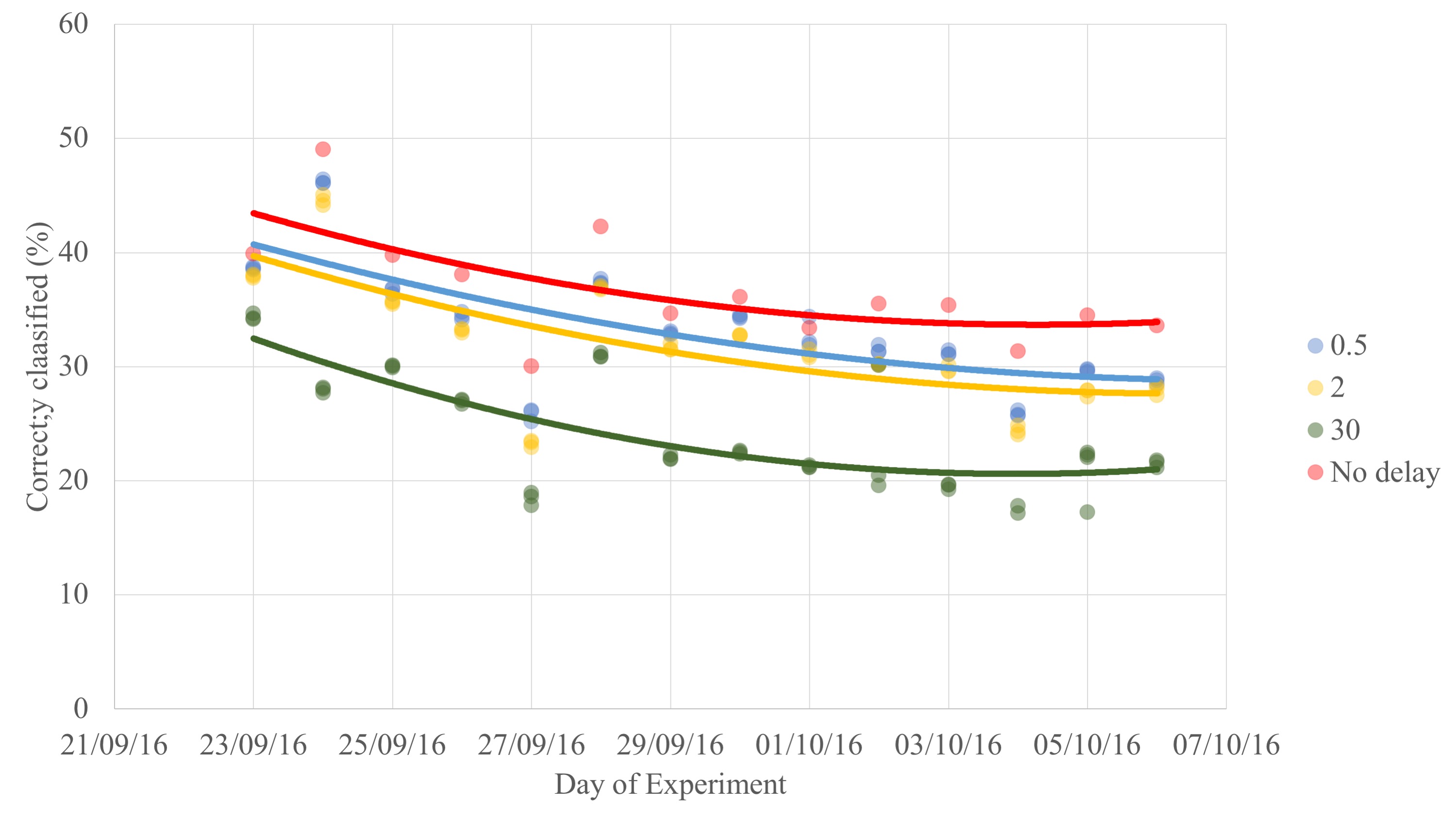} }
		%\subfloat[]{\label{fig:blind-delay} \includegraphics[width=8cm,height=7cm,keepaspectratio]{pic/blind-delay.jpg} }			
		\subfloat[]{\label{fig:blind-delay-day} \includegraphics[width=9cm,keepaspectratio]{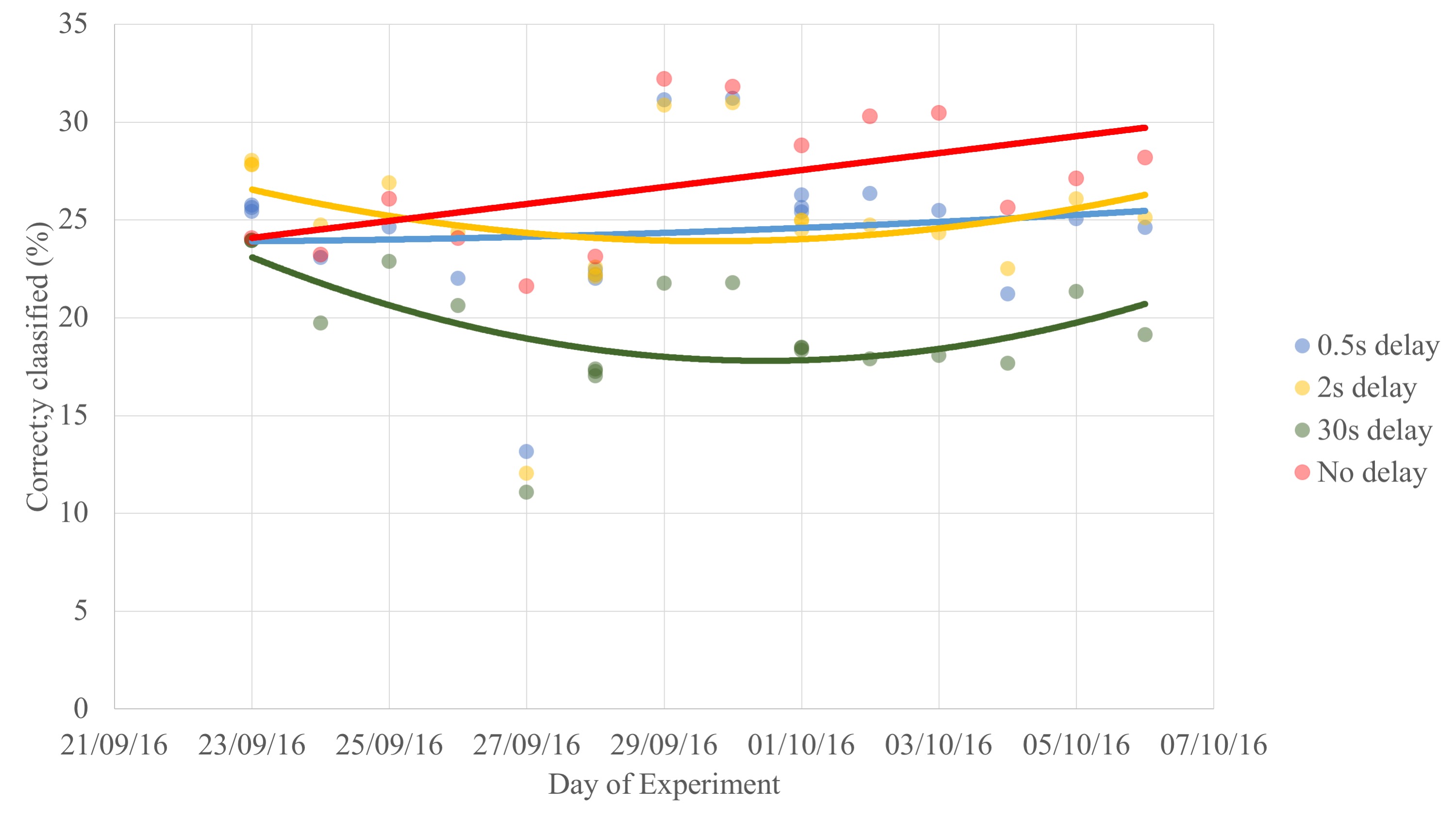} }
		%}
		\caption{The combined impact of multi-packet transactions and delayed for informed attacks (a) and blind attacks (b)}	\label{fig:multipacketledger-delayed}
		%\label{fig:combination-localIL-and-transactions}
	\end{center}
	
\end{figure*}
The maximum separation time for two consecutive packets from the same device varies across devices depending on their function. Combining multiple packets into a transaction conceals  short separation times which allowed us to build recognition models for individual devices.  Additionally,  it allows to encompass the whole peak of data, when there is one. The exact number of packets that are combined together depends on application and the total number of packets generated by the device. Figure~\ref{fig:multipacket} shows the results for informed and blind attacks. For informed attackers, multi-packet transactions decrease the classification accuracy by about 20\% on average. The attacker has clear understanding of the patterns of transactions and thus uses short separation times to discriminate a device from another. By consolidating multiple packets to a single transaction and thus removing the separation time, the rate of correct identification is reduced.  

The change is less obvious for blind attackers, but the results obtained after this operation are still lower than the ones we get with the baseline method. Recall that a blind attacker only uses a subset of actual devices in its training set, so it has lower visibility  to distinguish the devices, basing it only on higher separation times. Concealing the shortest separation times with multi-packet transactions has a greater impact on an informed attacker than on a blind attacker given the informed attacker's full visibility into the device type and their short separation times. There is significant variation in the differences between single and multi-packet transactions over different days in the simulation for blind attacks, as the performance depends significantly on the active devices on that day and whether these devices are in the blind attacker's training set.

\begin{figure*} 	% h = means that exaclty here
	\begin{center}
		%\subfloat[]{\label{fig:cloud-storage-pic} \includegraphics[width=7cm,height=7cm,keepaspectratio]{pics/storage.jpeg} }\hfill
		%	\mbox{
		%\subfloat[]{\label{fig:informed-delay} \includegraphics[width=8cm,height=7cm,keepaspectratio]{pic/informed-delay.jpg} }
		\subfloat[]{\label{fig:informed-delay-day} \includegraphics[width=9cm,keepaspectratio]{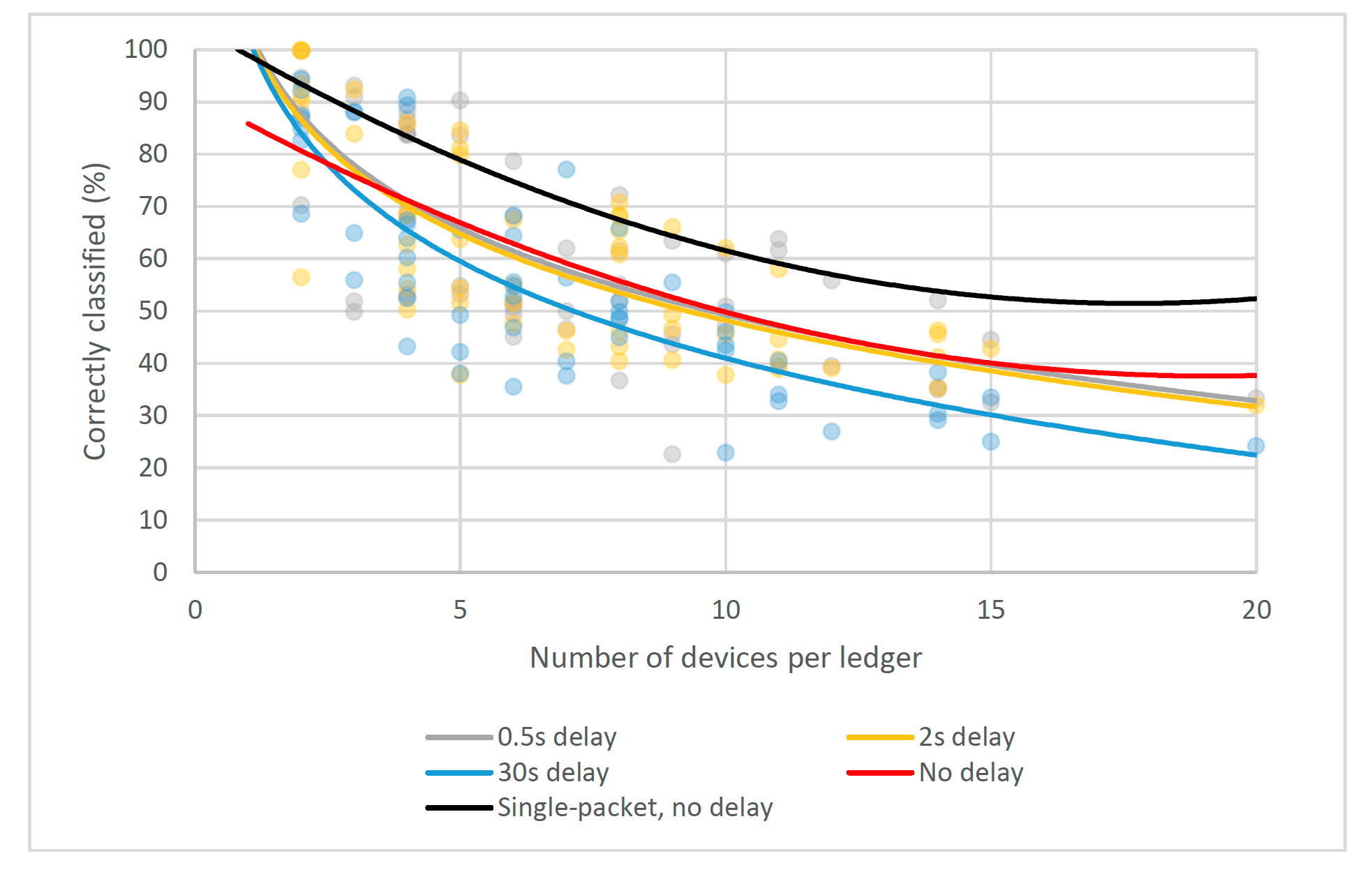} }
		%\subfloat[]{\label{fig:blind-delay} \includegraphics[width=8cm,height=7cm,keepaspectratio]{pic/blind-delay.jpg} }			
		\subfloat[]{\label{fig:blind-delay-day} \includegraphics[width=9cm,keepaspectratio]{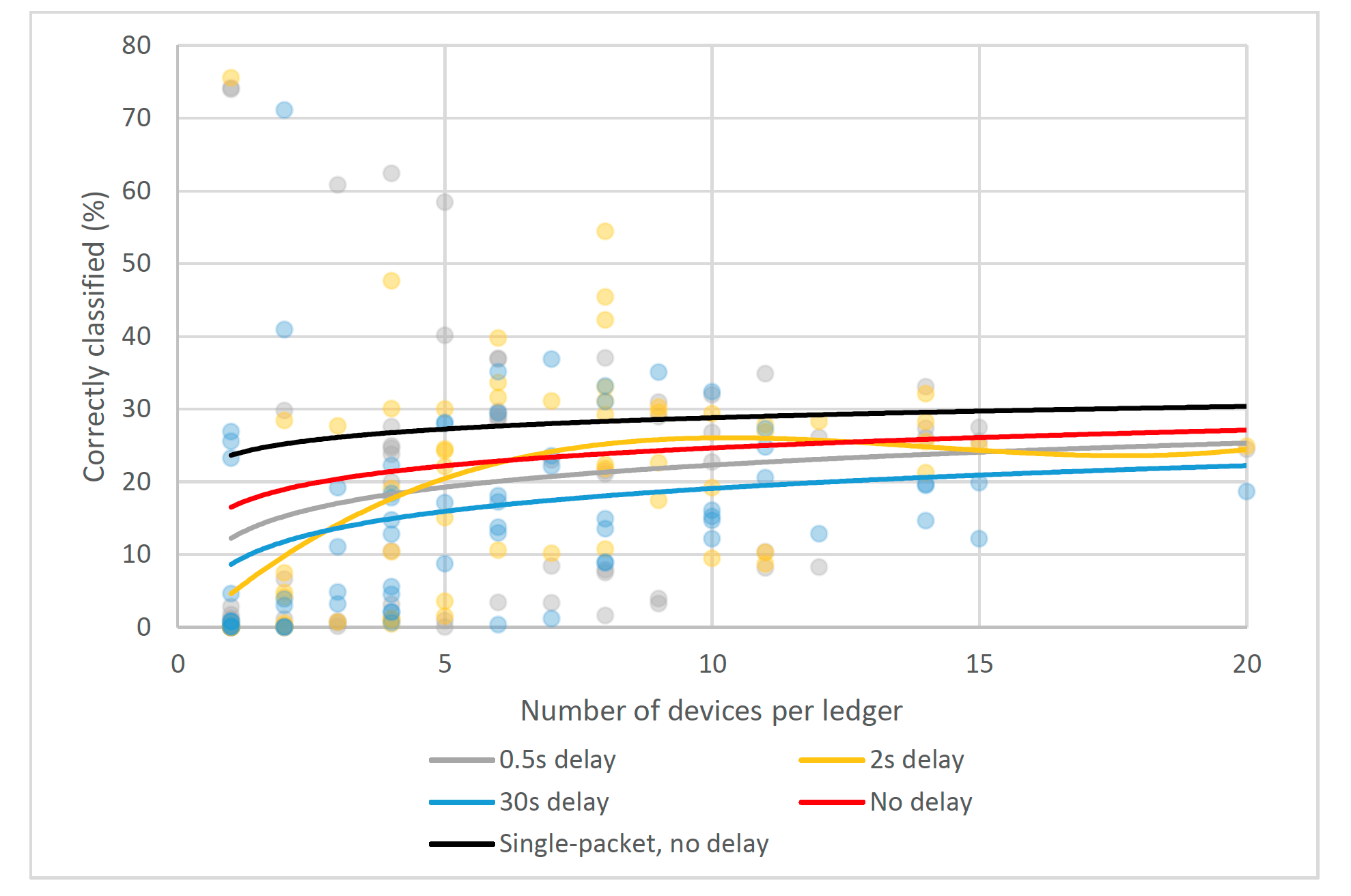} }
		%}
		\caption{The combined impact of all three methods for informed attacks (a) and blind attacks (b)}	\label{fig:combined}
		%\label{fig:combination-localIL-and-transactions}
	\end{center}
	
\end{figure*}

\subsection{Combined Timestamp Obfuscation}

We have seen that all the timestamp obfuscation methods, applied separately, increase resilience against device identification compared to the baseline case. Noting that the percentage of correctly classified transactions should be around 5\% for a strategy that randomly picks a device from the set (\,knowing that our experimental set is made of about 20 devices)\, we now evaluate the effects of combining our obfuscation measures. \newline
We first combine multi-packet transactions with multi-node ledgers, and report the results in Figure \ref{fig:multipacketledger}. For informed attackers though, the combination of multi-packet transactions with multi-node ledgers does improve the privacy by about 10\% all along, as the additional concealment of fine-grained temporal features reduces the value of the informed attacker's full visibility into device types. Combining multi-packet transactions with multi-node ledgers yields results extremely close (\,less than 5\% of difference)\, to the ones obtained with single-packet transactions, for blind attackers. Again, the reason is that blind attackers rely more on longer packet separation time and are less impacted by obfuscations affecting short separation times.  \newline

Next, we combine delayed transactions with multi-packet transactions, and the results are shown in Figure~\ref{fig:multipacketledger-delayed}. Delaying transactions alongside multi-packet transactions yields similar results as before: delaying transactions by up to 0.5 or 2 seconds gives similar results, and delaying transactions by up to 30 seconds significantly improves the privacy, reducing the classification accuracy by 10\% or more. However, if we compare these results to the ones obtained when using single-packet transactions, we surprisingly notice that the privacy is actually better with delayed single-packet transactions. One reason could be that, in our raw sample, the distribution of the transactions within the different devices is relatively balanced (with roughly comparable numbers of transactions per device). Multi-packet transactions can cause significant changes in the transaction distribution across devices, with some devices becoming more dominant and others becoming scarcely represented. This results in classifiers achieving high performance by simply guessing the more dominant devices all the time.

Finally, we compare the impact of combining all three timestamp obfuscation measures, shown in Figure~\ref{fig:combined}. For informed attackers, multi-packet transactions with no delay perform best for ledgers with 1-2 devices. The likely reason is that the delay has diminishing returns on timestamp obfuscation when both multi-packet transactions and multi-node ledgers are in place. We note though that results for 1-2 devices are more sensitive to the simulation's random selection of devices for each run, which brings in dependencies of the selected device(s)' temporal features. The combined approach of all three measures and a delay of 30 seconds performs best overall, achieving the lowest classification accuracy of 24\%. For blind attacks, there is a slight advantage for the combined approach with 2 second delays for ledgers of 1-2 nodes, while again overall the combined approach with a 30 second delay achieves the attack success rate at 19\%.

\section{Discussion and Conclusion}\label{sec:discussions}
In this paper, we analyzed device classification in IoT-based blockchain which is, to the best of our knowledge, the first to provide such analysis.  We show that device identification in blockchain-based IoT introduces privacy concerns. Unlike conventional IoT device classification methods where the identifier requires to have physical access to the network, in blockchain-based IoT any entity can identify devices independent of the location of the entity. We used a smart home setting as a representative case study of the IoT.  We applied machine learning algorithms on the blockchain to classify the devices.  The results have demonstrated that  the attacks can have up to 90\% accuracy in  classifying    type and number of devices in a smart home. To reduce the success rate of device classification, we proposed three timestamp obfuscation methods such as combining multiple packets into a single transaction, merging ledgers of multiple packets, and randomly delaying transactions. The proposed timestamp obfuscation methods can reduce the success rate to below 30\%. \par
By monitoring the frequency of the transactions generated by IoT devices, e.g., motion sensors or smart lock, the attacker can identify the hours that the smart home is occupied or is empty which may lead to further security risks, e.g., robbery.  Additionally, the attacker can compromise the privacy of the user as it can monitor the 	communications of the IoT devices, stored as transactions in the blockchain, that   reveal information about the user activities.   As  future work,  we plan to study device identification and user deanonymization in other  applications which includes energy trading and smart cities. Additionally, we plan to study the impact of the training set on the informed attack.

\bibliographystyle{IEEEtran}
\bibliography{conference_041818}

\end{document}